\documentclass[
aip,
 amsmath,amssymb,
 reprint,%
]{revtex4-2}

\usepackage{graphicx}
\usepackage{dcolumn}
\usepackage{bm}
\usepackage{hyperref}

\usepackage[utf8]{inputenc}
\usepackage[T1]{fontenc}
\usepackage{mathptmx}
\usepackage{etoolbox}
\usepackage{xcolor}

\newcommand{\Google}{\affiliation{Google Quantum AI, Goleta, California 93117, USA}}

\begin{document}
\begin{title}

\title{Josephson parametric circulator with same-frequency signal ports, 200~MHz bandwidth, and high dynamic range}

\author{Randy Kwende}\Google
\affiliation{Department of Electrical and Computer Engineering, University of Massachusetts, Amherst, Massachusetts 01003, USA}
\author{Theodore White}\Google
\author{Ofer Naaman}\email{ofernaaman@google.com}\Google

\date{\today}

\begin{abstract}
We demonstrate a 3-port Josephson parametric circulator, matched to $50\,\Omega$ using second order Chebyshev networks. The device notably operates with two of its signal ports at the same frequency and uses only two out-of-phase pumps at a single frequency. As a consequence, when operated as an isolator it does not require phase coherence between the pumps and the signal, thus simplifying the requirements for its integration into standard dispersive qubit readout setups. The device utilizes parametric couplers based on a balanced bridge of rf-SQUID arrays, which offer purely parametric coupling and high dynamic range. We characterize the device by measuring its full $3\times3$ S-matrix as a function of frequency and the relative phase between the two pumps. We find up to 15~dB nonreciprocity over a 200~MHz signal band, port match better than 10~dB, low insertion loss of 0.6~dB, and saturation power exceeding $-80$~dBm.
\end{abstract}

\maketitle
\end{title}

Microwave circulators and isolators are indispensable components of the readout subsystem in superconducting quantum processors, enabling signal routing and high efficiency measurement while protecting the qubits from amplifier and thermal noise. A typical dispersive readout setup employs several ferrite circulators\cite{arute2019quantum, krinner2022realizing} between the qubit chip and the first amplification stage in each measurement channel. Even when the readout of several qubits is multiplexed on a single measurement channel, the ferrite circulators occupy a significant fraction of the available space on the mixing chamber plate of a typical dilution refrigerator. For this reason, a long standing research objective on the path to miniaturizing the readout subsystem has been to find a microelectronic substitute for the ferrite circulator.

Circulation relies on two essential phenomena: interference, and nonreciprocal scattering\textemdash signals traveling in opposite directions experience different phase shifts. In ferrite circulators\cite{bosma1968general, fay1965operation}, nonreciprocity is a result of the anisotropic magnetic permeability of the ferrite material. In microelectronic circulators, nonreciprocity is realized using temporal modulation of reactances or couplings in the circuit:\cite{ranzani2019circulators} this includes spatio-temporal modulation of resonator frequencies on a ring\cite{estep2014magnetic, kerckhoff2015chip}, frequency mixing\cite{reiskarimian2016magnetic, ruffino2020wideband, chapman2017widely}, and parametric conversion\cite{sliwa2015reconfigurable, lecocq2017nonreciprocal, ranzani2017wideband, abdo2017gyrator, alvarez2019coupling}. 

Parametric nonreciprocity based on superconducting Josephson circuits is particularly suitable for quantum computing applications\cite{ranzani2019circulators} because in principle it can operate with very low intrinsic dissipation and low insertion loss. However, Josephson parametric circulators have so far been limited to a narrow band resonant response and low saturation powers. These devices are typically designed with a distinct resonant frequency at each port, and require multiple phase-coherent pump tones at different frequencies\cite{sliwa2015reconfigurable, lecocq2017nonreciprocal} to activate multiple parametric conversion processes.

Here, we describe a Josephson parametric circulator based on a core of three coupled resonant modes, similar to Refs.~\onlinecite{sliwa2015reconfigurable, lecocq2017nonreciprocal}. However, rather than using three distinct frequencies, two of the resonant modes are operated at the same frequency and are coupled passively to each other. The third mode is at a different frequency, and is coupled to the other two via parametric conversion processes driven by two pumps that have the same frequency but are out of phase. As a result, the transmission between the two same-frequency ports is independent of the relative phase between the signal and the pumps. Additionally, the two pumps could in principle be supplied from two phase-shifted copies of a single pump tone from a single generator. This significantly reduces the barriers to the device's integration into a typical readout system. Like ferrite circulators, Josephson parametric circulators are resonant circuits. Ferrite circulators are made broad-band by the addition of stepped-impedance matching networks on each of their ports\cite{helszajn2008stripline}. Similar techniques\cite{anderson1967analysis} can be used to increase the bandwidth of parametric circulators\textemdash our device is matched to the $50\,\Omega$ ports with 2-pole Chebyshev networks\cite{naaman2022synthesis} (designed for 250~MHz bandwidth). The device's active elements are built using rf-SQUID array balanced-bridges\cite{white2022readout, naaman2017josephson} for high dynamic range performance. 

\begin{figure*}[ht]
\includegraphics[width=\textwidth]{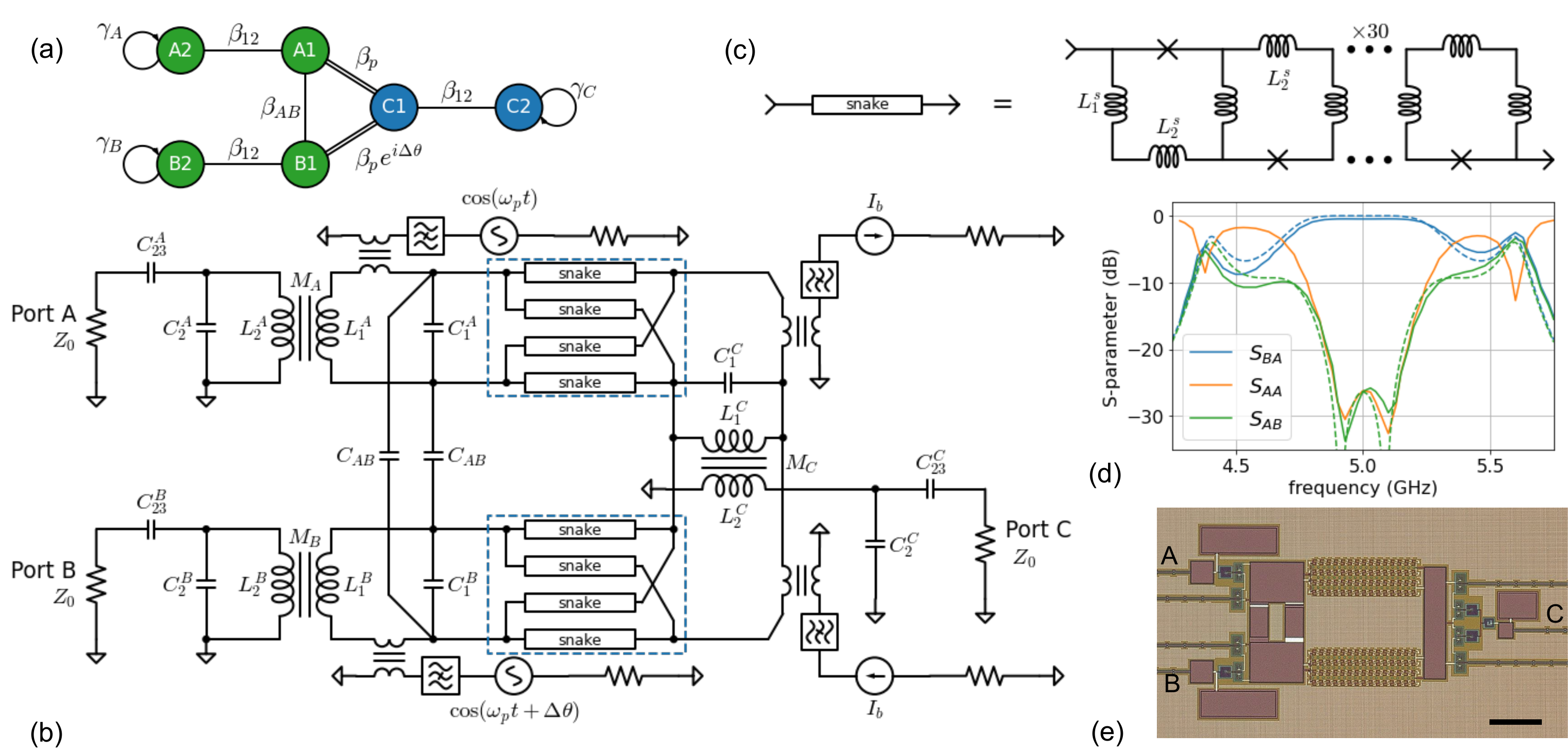}
\caption{\label{fig:circuit} (a) Coupled-mode graph of the 2nd-order matched circulator. The $A$- and $B$-modes are resonant at 5 GHz (green), and the C-modes are resonant at 7 GHz (blue). Modes $A1$ and $B1$ are coupled passively with $\beta_{AB}$, while mode $C1$ is coupled parametrically to $A1$ and $B1$ with $|\beta_p|=\beta_{AB}=0.5$.  (b) Simplified circuit schematic of the device. Crossing wires are only connected if a `solder dot' exists at their intersection. For the $A$ and $B$ modes, $L_1^A=L_1^B = 400$~pH, $C_1^A=C_1^B=4.36$~pF, $L_2^A=L_2^B=405$~pH, $C_2^A=C_2^B=2.22$~pF, $M_A=M_B=65$~pH, and $C_{23}^A=C_{23}^B=0.57$~pF. For the $C$ modes $L_1^C=200$~pH, $C_1^C=3.72$~pF, $L_2^C=202$~pH, $C_2^C=1.1$~pF, $M_C=57$~pH, and $C_{23}^C=0.246$~pF. The coupling $\beta_{AB}$ between modes $A1$ and $B1$ is provided by capacitors $C_{AB}=1.09$~pF. (c) Schematic of the snake rf-SQUID array\cite{white2022readout, naaman2017josephson}. (d) Harmonic balance (solid) and coupled-mode (dashed) simulations of the parametric circulator. (e) Optical micrograph of the device, scale bar is $120\,\mu$m. The signal ports are labeled on the figure. Pumps are fed from the left center, flux biases are applied from the right.}
\end{figure*}

Figure~\ref{fig:circuit}(a) shows the coupled-mode graph\cite{naaman2022synthesis} of the device. The $A$ and $B$ modes (green) have a resonance frequency of $\omega_A=\omega_B=2\pi\times5$~GHz, and the $C$ modes (blue) resonate at $\omega_C=2\pi\times 7$~GHz. Modes $A1$, $B1$, and $C1$ form the circulator core\cite{sliwa2015reconfigurable, lecocq2017nonreciprocal}, and the rest are added to provide impedance matching with a 2-pole Chebyshev response\cite{naaman2022synthesis} using the 0.01~dB ripple filter prototype coefficients\cite{pozar2009microwave}
\begin{equation}\label{eq:proto}
    \left\{g_0,\dots,g_3\right\}=\left\{1.0,\,0.449,\,0.408,\,1.101\right\},
\end{equation}
where $g_0$ represents the effective conductance of the circulator core\cite{anderson1967analysis}, and $g_3$ represents the $50\,\Omega$ load. Modes $A2$, $B2$, and $C2$ are coupled to $50\,\Omega$ ports with a dissipation rate of $\gamma_0\equiv\gamma_A=\gamma_B=\gamma_C$, which can be calculated from the prototype Eq.~(\ref{eq:proto}), $\gamma_0=\Delta\omega/g_2g_3 = 2\pi\times557$~MHz, where $\Delta\omega/2\pi=250$~MHz is the design bandwidth of the network. The reduced passive coupling rate between the filter modes is given by $\beta_{12}=\Delta\omega/2\gamma_0\sqrt{g_1g_2} = 0.525$, and the core modes are coupled pair-wise with a reduced rate $\beta_{AB}=|\beta_p|=0.5$, where $\beta_{AB}$ is a passive coupling, and $\beta_p$ is a parametric coupling driven at $\omega_p = \omega_C-\omega_A$, whose strength depends on the pump amplitude. The parametric coupling on the edge $B1-C1$ has a phase difference of $\Delta\theta$ relative to that on the $A1-C1$ edge, and the phase difference, ideally $\pm\pi/2$, determines the direction of circulation.

Figure~\ref{fig:circuit}(b) shows a simplified circuit schematic of the device. The nonlinear elements are indicated in the figure as eight boxes labeled `snake', appearing in two groups (highlighted in blue in the figure): each of these is a linear array of 30 interleaved rf-SQUIDs\cite{white2022readout, naaman2017josephson} [see Fig.~\ref{fig:circuit}(c)], whose construction, biasing method, and theory of operation were described in detail in Ref.~\onlinecite{white2022readout}. In each group, the four snake arrays are arranged in a balanced bridge configuration forming a parametric coupler, and two of these couplers are used in the circuit to implement the $A1-C1$ and $B1-C1$ edges. The couplers are flux biased to set their operating point (where the inductance of each snake array is $L_\mathrm{snake}=427$~pH), and are flux pumped out of phase at the same frequency $\omega_p$. Both pump and bias lines include on-chip notch filters that prevent signals in the $5-7$~GHz range from propagating out of the device. Modes $A1$ and $B1$ are constructed using inductances $L_1^A=L_1^B$ and capacitances $C_1^A=C_1^B$. Modes $A2$ and $B2$ have $L_2^A=L_2^B$ and $C_2^A=C_2^B$. Mode $A2$ ($B2$) is inductively coupled to $A1$ ($B1$) via mutual inductance $M_{A(B)}$, and is capacitively coupled to the $50\,\Omega$ port with $C_{23}^{A(B)}$. The $C_{1(2)}$ modes comprise of $L_{1(2)}^C$ and $C_{1(2)}^C$ and couplings $M_C$ and $C_{23}^C$. When the bridges are balanced, they present each of the core modes with an additional inductance of $L_\mathrm{snake}$ in parallel with their primary inductance. Finally, the coupling $\beta_{AB}$ between modes $A1$ and $B1$ is provided by capacitors $C_{AB}$. Component values are listed in Fig.~\ref{fig:circuit} caption.

The balanced-bridge design of the couplers ideally ensures zero passive coupling between the modes, and the coupling is therefore ideally purely parametric. Either of the dc flux bias or the pumps, when applied separately, change the inductances of the snake arrays, but do so equally in all four arrays keeping the bridge balanced by symmetry. Only the combination of the bias and pump currents applied together can unbalance the bridge, providing a positive coupling in one half-cycle of the pump, and a negative coupling in the second half-cycle\cite{naaman2017josephson}. The time-average (passive) coupling is therefore zero, and the coupling has only an ac (parametric) component. Each of the bridges can then be modeled as a mutual inductance $M(t)=M_0+\delta M\cos(\omega_p t+\theta_p)$, where $M_0=0$, $\delta M$ is controlled by the pump amplitude and the flux bias, $\omega_p$ is the pump frequency and $\theta_p$ is the pump phase. To facilitate a transition from the single-ended topology of the ports to the balanced topology of the core resonators, we chose to use mutual inductive coupling ($M_A$, $M_B$, $M_C$) between the core resonators ($A_1$, $B_1$, $C_1$) and the matching resonators ($A_2$, $B_2$, $C_2$).

Figure~\ref{fig:circuit}(c) shows a schematic of the snake rf-SQUID array. The meandering inductive spine of the snake has alternating $L_1^s=2.6$~pH and $L_2^s=8.0$~pH, and the junctions have nominal critical current of $I_c=16\,\mu$A. Each array contains 30 junctions. 

Figure~\ref{fig:circuit}(d) shows harmonic balance simulations of the circuit in panel (b) in solid lines, with the component values reported in the caption, for right-handed circulation with $\Delta\theta=-77^\circ$ (left handed circulation, not shown, is found at $\Delta\theta=80^\circ$). The dashed lines are showing calculated S-parameters\cite{naaman2022synthesis} based on the coupled-mode graph in panel (a). There is excellent agreement between the two simulation methods, which predict 25~dB of reverse isolation and port match with a 250~MHz equiripple bandwidth as-designed.

Figure~\ref{fig:circuit}(e) shows an optical micrograph of the fabricated device, built in a three-layer aluminum process with trilayer Al/AlO$_x$/Al Josephson junctions. The snake balanced bridges are visible in the center of the image. The $A$ and $B$ ports are labeled on the left, and port $C$ is on the right. The image also shows the pump (left center) and flux bias (right) lines.

\begin{figure}
\includegraphics[width=\columnwidth]{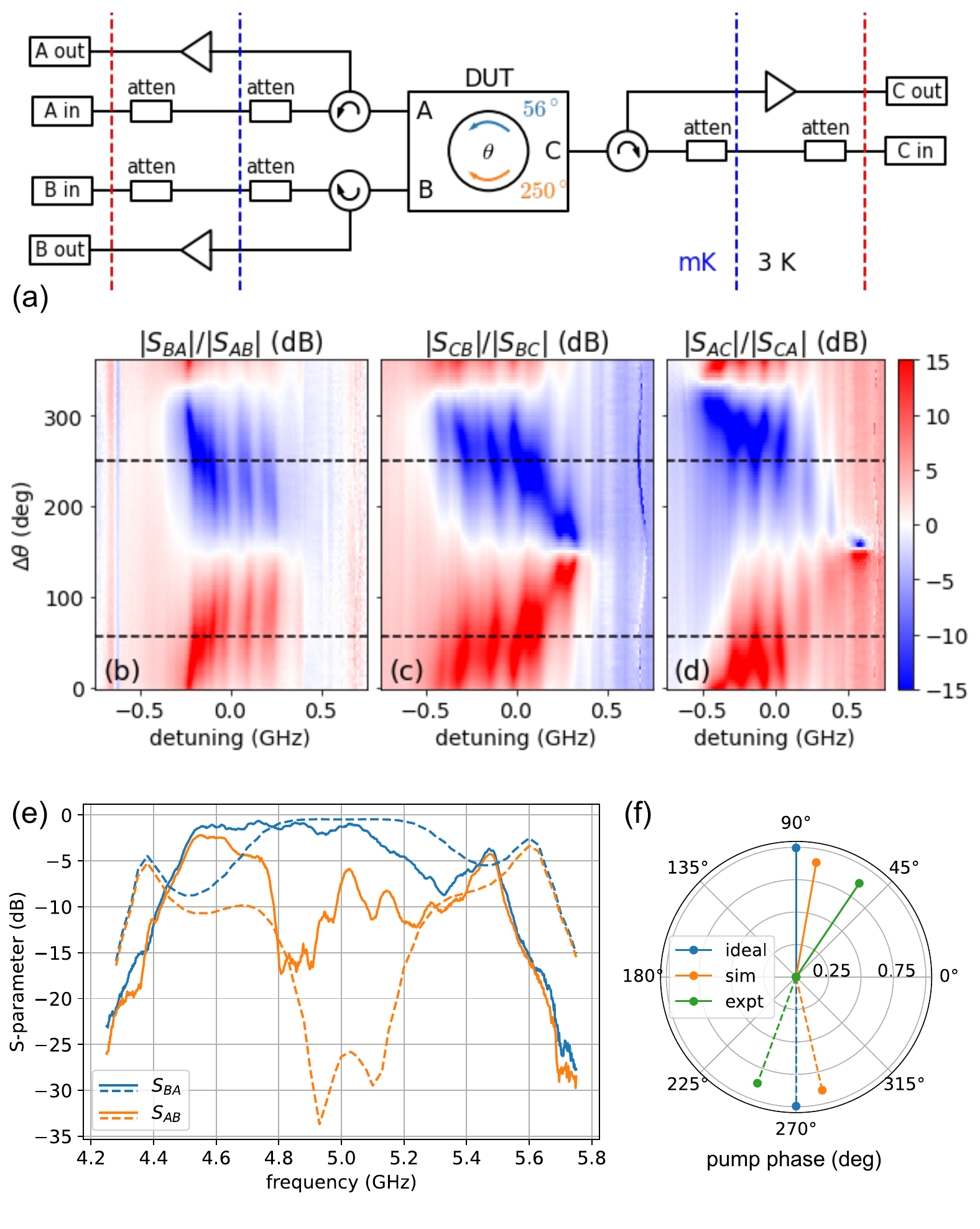}
\caption{\label{fig:nonrecip} (a) Schematic of the measurement setup. (b)-(d) The nonreciprocity of the device (in dB), defined as the ratio of the forward and reverse S-parameters, plotted vs signal frequency and relative pump phase $\Delta\theta$. (b) Nonreciprocity between the same-frequency input and output ports, the x-axis indicates detuning from 5 GHz. (c) and (d) are nonreciprocities involving the frequency converting 7~GHz port $C$. The x-axis here represent detuning from either 5~GHz or 7~GHz. Horizontal dashed lines represent the pump phase settings in the right-handed circulation direction, $\Delta\theta=56^\circ$, and the left-handed direction, $\Delta\theta=250^\circ$. (e) $S_{BA}$ and $S_{AB}$ (solid), compared to ideal circuit simulation (dashed). (f) Relative pump phase in RH (solid) and LH (dashed) circulation, for the ideal circuit (blue), simulation (orange), and data (green). The length of each vector represents the corresponding insertion loss, $10^{-\mathrm{IL}/10}$.}
\end{figure}

The device was packaged in a magnetically shielded enclosure and measured in a dilution refrigerator with $T<20$~mK. The experimental setup is shown schematically in Fig.~\ref{fig:nonrecip}(a), allowing measurement of all nine S-parameters of the 3-port device. Input lines are attenuated by 20~dB at the 3~K stage, and by an additional 40~dB at the mixing chamber of the fridge. Accounting for additional cable losses, the total input attenuation is estimated around $65\pm5$~dB. Out-going signals (either reflected at the ports or transmitted through the device) are routed via ferrite circulators to HEMT amplifiers at 3~K followed by additional low noise amplifiers at room temperature (not shown). DC flux bias was applied using two room temperature voltage sources through 20~dB attenuators at 3~K and low-pass and IR filters at the mixing chamber. The device was pumped at $\omega_p/2\pi=2.004$~GHz using two phase-coherent channels of a dual signal generator (Berkeley Nucleonics BNC-855) with an adjustable relative phase. Both pumps powers were $-13.6$~dBm at room temperature, and fed to the device via $30$~dB of attenuation. S-parameters were measured using a Keysight PNA-L vector network analyzer (VNA) equipped with a scalar frequency conversion measurement option. For these experiments, we did not perform cryogenic calibration of the VNA to the reference plane of the device, instead we rely on measurement of the through transmission in each measurement path, bypassing the device with a coaxial cable in a separate cooldown. 

We measured all S-parameters as a function of frequency for varying relative phase $\Delta\theta$ of the pumps. In Figure~\ref{fig:nonrecip}(b)-(d), we plot the nonreciprocity\textemdash the difference, in dB, between forward ($A\rightarrow B\rightarrow C\rightarrow A$) and reverse ($C\rightarrow B\rightarrow A\rightarrow C$) transmission\textemdash for all port pairs as a function for relative pump phase $\Delta\theta$. We see up to 15~dB nonreciprocity with clear $\Delta\theta$ dependence. When $\Delta\theta=56^\circ$, circulation is `right handed' (RH) (as defined by Fig.~\ref{fig:nonrecip}) and when $\Delta\theta=250^\circ$ the circulation is left-handed (LH). Aside from an overall phase offset due to differences in signal path lengths, the phase difference between RH and LH circulation deviates from the ideal $180^\circ$. This is also observed in simulations, and is likely due to the inherent frequency dependence of all coupling elements or frequency misalignment between the modes [see panel~(f)]. Panel~(b) represents the nonreciprocal scattering between the two same-frequency ports $A$ and $B$, with maximum nonreciprocity over a 200~MHz band. Panels (c) and (d) demonstrate directional scattering involving frequency conversion into port $C$. Throughout the band of the device, we see regions of reduced nonreciprocity, appearing as pale vertical stripes in the figure, that are weakly dependent on $\Delta\theta$. These could be the result of parasitic passive coupling between the ports; they reduce the effective useful bandwidth of the device. Fig.~\ref{fig:nonrecip}(e) shows the measured $S_{BA}$ and $S_{AB}$ for RH circulation (normalized to a through, solid) in comparison to a simulation of the ideal circuit (dashed). We observe a slight shift of the band, and overall reduced bandwidth and isolation compared to the design target. Fig.~\ref{fig:nonrecip}(f) depicts, on a polar plot, the relative pump phase in the ideal case (blue), circuit simulation (orange), and the experiment (green), for RH (solid) and LH (dashed) circulation. The length of each vector represents the corresponding insertion loss (IL), $10^{-\mathrm{IL}/10}$. 

\begin{figure}
\includegraphics[width=\columnwidth]{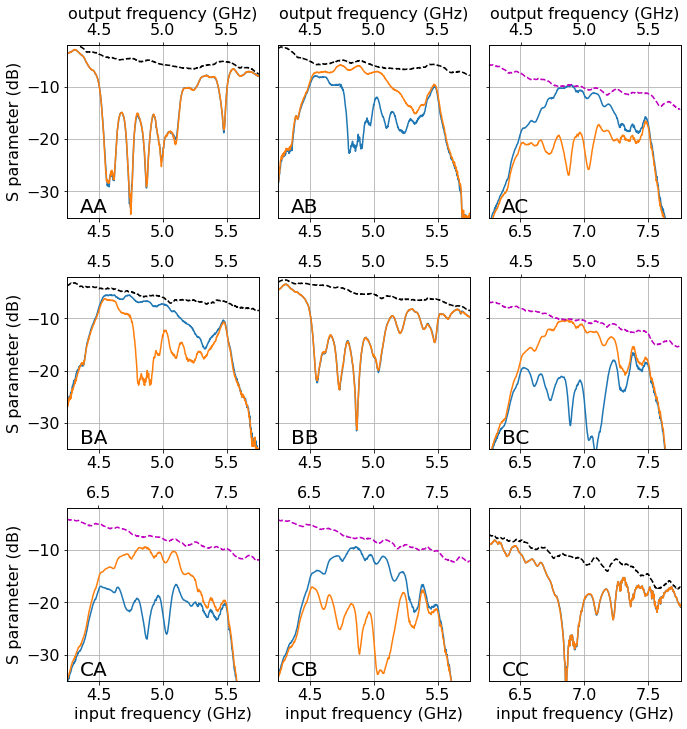}
\caption{\label{fig:sparams} The $3\times 3$ S-matrix of the circulator vs frequency at input power $\approx -105$~dBm. Each panel is labeled with the corresponding S-parameter, where the first index is the output port and the second index is the input port. In each panel, the input frequency is shown on the lower x-axis, and the output frequency is shown on the top x-axis. The blue trace in each of the panels was measured for relative pump phase of $\Delta\theta = 56^\circ$, and the orange curve was measured for $250^\circ$. Black dashed lines: measured through transmission. Magenta dashed lines: estimated through transmission as the average (in dB) of data in the input and output frequency bands.}
\end{figure}

Figure~\ref{fig:sparams} shows the raw $3\times 3$ S-matrix of the device. Each panel shows two traces: in blue we plot the S-parameter for $\Delta\theta=56^\circ$ (right-handed circulation), and in orange we plot the same S-parameter for $\Delta\theta=250^\circ$ (left-handed circulation). In the figure, each panel has two x-axes: the lower one represents the input frequency, and the upper one represent the output frequency. Black dashed lines in the figure represent through transmission for the respective signal paths, measured by bypassing the device and its package in a separate cooldown. Magenta dashed lines are baseline through for the frequency converting matrix elements; these cannot be measured directly, so they are estimated from the geometric mean (average dB) of through measurements at the input and output frequency bands. 

The diagonal elements of Fig.~\ref{fig:sparams}, showing $S_{AA}$, $S_{BB}$, and $S_{CC}$, indicate that all ports are matched with return loss of better than $10$~dB around the center frequencies of the $A$ and $B$ modes (4.86~GHz) and the $C$ mode (6.9~GHz), and their responses are identical in the two $\Delta\theta$ settings. 

Focusing on the right-handed circulator setting with $\Delta\theta=56^\circ$ (blue traces), we see from $S_{AB}$ that signals entering port $B$ are prevented from reaching port $A$, with 15~dB isolation at 4.86~GHz and a bandwidth of 200~MHz. The $S_{CB}$ trace shows that indeed signals entering the device from port $B$ are diverted instead to emerge upconverted from port $C$. At the same time, $S_{BA}$ shows full transmission in the forward direction of signals from port $A$ to port $B$, with a minimum insertion loss of $0.6$~dB. Signals entering port $C$ scatter downconverted into port $A$ ($S_{AC}$) but not port $B$ ($S_{BC}$). The roles reverse for $\Delta\theta=250^\circ$ (orange). Overall, the scattering matrix of the device at $\Delta\theta=56^\circ$ approximates the ideal right handed circulator S-matrix $\mathbf{S}_\mathrm{RH}$, and at $\Delta\theta=250^\circ$ the device approximates the ideal left-handed circulator $\mathbf{S}_\mathrm{LH}$:
\begin{equation}
    \mathbf{S}_\mathrm{RH} =
    \begin{bmatrix}
    0 & 0 & 1 \\
    1 & 0 & 0 \\
    0 & 1 & 0
    \end{bmatrix},\;\;\;
    \mathbf{S}_\mathrm{LH} =
    \begin{bmatrix}
    0 & 1 & 0 \\
    0 & 0 & 1 \\
    1 & 0 & 0
    \end{bmatrix}.
\end{equation}

\begin{figure}
\includegraphics[width=2.8in]{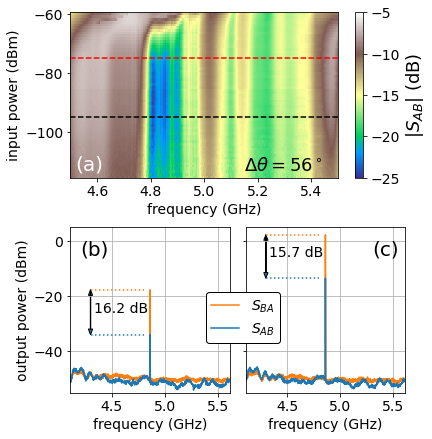}
\caption{\label{fig:power_spect} (a) $S_{AB}$ vs frequency as a function of input power for $\Delta\theta=56^\circ$. The y-axis indicates the estimated power at the chip: power applied at room temperature ranged from $-50$~dBm to $+5$~dBm, and the attenuation in the signal path is $65\pm5$~dB. Saturation is observed for input powers around $-70$~dBm. Black and red dashed lines indicate the power levels used in (b) and (c) respectively. (b) Spectrum analyzer traces in the forward $S_{BA}$ (orange) and reverse $S_{AB}$ (blue) directions with an $\omega/2\pi=4.86$~GHz, $-95$~dBm input tone. (c) Spectra for a higher signal power, $-75$~dBm. In both (b) and (c) the y-axis is the the output power measured at room temperature, and includes gain from multiple stages of amplification.}
\end{figure}

The insertion loss of the device can be attributed to several factors, including mismatch losses at the ports, dissipative losses in the package, dielectric losses in the device itself, and residual coupling of signal currents in the bridge to the $50\,\Omega$ environment through the bias and pump lines, despite our attempt to mitigate this loss channel by use of notch filters.

To be useful in multiplexed readout of quantum processors, the parametric circulator must have sufficiently high dynamic range to avoid saturation and intermodulation distortion at the total power of all simultaneous qubit readout tones. For this reason, our design uses rf-SQUID `snake' arrays [Figure~\ref{fig:circuit}(c)] similar to those used in the high dynamic range parametric amplifiers described in Ref.~\onlinecite{white2022readout}. Figure~\ref{fig:power_spect}(a) shows the isolation $S_{AB}$ for $\Delta\theta=56^\circ$ as a function of signal frequency and input power. The input power here is referenced to the circulator's input based on an estimated $65$~dB of attenuation in the fridge [Fig.~\ref{fig:nonrecip}(a)]. We see that there are no significant changes to the S-parameters up to at least $-80$~dBm of input signal power, but a clear degradation of the performance above $-70$~dBm. This is comparable to the measured output saturation power of parametric amplifiers\cite{white2022readout} based on similar rf-SQUID arrays, and is significantly higher than the typical $-120$~dBm to $-110$~dBm total power used in multiplexed qubit readout\cite{khezri2022measurement}.

Figure~\ref{fig:power_spect}(b) shows the output spectra in response to an input tone with $P=-95$~dBm [black dashed line in panel (a)] at 4.86~GHz in both the forward ($A\rightarrow B$, orange) and isolated ($B\rightarrow A$, blue) directions. Panel (c) shows similar spectra but with $P=-75$~dBm [dashed red line in Fig.~\ref{fig:power_spect}(a)]. We see that the spectra are free of spurious signals, with no excess noise even at the higher signal powers. We note that multiples of the pump frequency can appear at the device's output ports when these harmonics fall within the pass band of the device, and can even cause pump-signal intermodulation products to appear in the band. For example, we have tuned the device to a different operating point with 2.3~GHz pumps, and the pump's second harmonic at 4.6~GHz was clearly visible in the output spectrum. This pump leakage into the signal mode is not surprising since, as can be seen in Fig.~\ref{fig:circuit}(b), the pump and signal share the same current path through the core modes' primary inductance. An improved design should attempt to decouple the pump and signal current paths, for example by symmetry.

To conclude, a useful microelectronic circulator should have low power dissipation, low insertion loss, high directivity, high saturation power, and wide bandwidth. It should also avoid imposing additional complications and new requirements on system integration, such as multiple pump frequencies, phase coherence between pumps and signals, and frequency translated outputs. With this requirement list in mind, we designed, built, and experimentally demonstrated a Josephson parametric circulator with high saturation power and a bandwidth of 200~MHz as controlled by the design of its matching network, with better than 10 dB port match, low insertion loss of 0.6~dB, and 15~dB nonreciprocity in the band. The device requires only single frequency pumps driving two couplers with a fixed phase relation\textemdash the pumps can in principle be derived from a single generator using a splitter and a phase shifter. Alternatively, the direction of circulation could be controlled dynamically, e.g.~for signal routing, by the phasing of the pumps\cite{sliwa2015reconfigurable}. Two of the circulator ports operate at the same frequency, simplifying its integration, especially when the device is used as an isolator where the third (frequency converting) port can be terminated. The use of balanced couplers greatly simplifies the microwave design of these devices, since they provide purely parametric coupling with minimal passive crosstalk between the matching networks. 

State-of-the-art cryogenic ferrite circulators can achieve 0.2~dB insertion loss and better than 20~dB port match and isolation over a 4~GHz band. Further improvements in Josephson parametric circulators beyond the prototype we presented here will be required for them to perform competitively with ferrites, specifically better targeting of component values in fabrication and mitigation of parasitic coupling. To what extent can microelectronic circulators replace ferrite ones in the qubit readout chain? The methods we have demonstrated here help elevate this question to a discussion about systems optimization.

\begin{acknowledgements}
We are grateful to the Google Quantum AI team for building, operating, and maintaining software and hardware infrastructure used in this work. We thank D. Sank and J. Bardin for comments on the manuscript.
\end{acknowledgements}

%

\clearpage
\end{document}